\documentclass{appolb}
\usepackage{graphicx}
\usepackage{amsmath}
\pdfoutput=1

\begin{document}
\title{Dark Forces Searches at KLOE-2%
}
\author{Elena P\'erez del R\'io \\ on behalf of the KLOE-2 Collaboration
\address{Laboratori Nazionali di Fracati dell'INFN \\ Via E. Fermi 40, 00044 Frascati, Italy}
\\
}
\maketitle
\begin{abstract}
Direct searches of dark matter are performed at accelerator facilities. The existence of a new vector boson has been postulated in different scenarios where in the most basic scheme the coupling to the SM can be achieved via a kinetic mixing term due to the U boson. The KLOE experiment at DA$\phi$NE searched for the U boson both in Dalitz decays of the $\phi$ meson and in continuum events. For all of these searches an upper limit for the U boson coupling $\epsilon^{2}$ has been established in the mass range $50 \, \text{MeV} < m_U < 1000\,\text{MeV}$. A summary of the different models and searches along with results are presented.
\end{abstract}
  
\section{Introduction}
The Standard Model (SM), although been the most complete theoretical framework at the present, does not provide a definitive model of all elementary particles.
In particular, recent observations as the $511$ keV gamma-ray signal from the galactic center~\cite{integral},the CoGeNT results~\cite{cogent},the DAMA/LIBRA annual modulation~\cite{dama,libra}, the total $e^+e^-$ flux~\cite{atic, hess1, hess2, fermi} and the muon magnetic discrepancy $a_{\mu}$ serve are examples of possible physics beyond the SM.
Extensions of the SM~\cite{ext1, ext2, ext3, ext4,ext5} claim to explain the afore-mention anomalies by dark matter models, with a Weakly Interacting Massive Particle (WIMP) belonging to a secluded gauge sector. The new gauge interaction would be mediated by a new vector gauge boson, the U boson or dark photon, which could interact with the photon via a kinetic mixing term,
\begin{equation}
\mathcal{L}_{mix} = - {\epsilon \over 2} F^{EM}_{\mu\nu}F^{\mu\nu}_{DM}
\end{equation}
where the parameter, $\epsilon$, represents the mixing strength and it is defined as the ratio of the dark to the SM electroweak coupling, $\alpha_{D}/\alpha_{EM}$.
A U boson, with mass of $\mathcal{O}(1 \text{GeV})$ and $\epsilon$ in the range of $10^{-2} - 10^{-7}$, could be observed in $e^+e^-$ colliders via different processes: $e^+e^- \rightarrow U \gamma$, $V \rightarrow P\gamma$ decays, where V and P are vector and pseudoscalar mesons, and $e^+e^- \rightarrow h'U$, where $h'$ is a Higgs-like particle responsible for the breaking of the hidden symmetry. 
On this basis, the KLOE experiment has performed several searches, which are reported.

\section{The KLOE detector at DA$\phi$NE}
The KLOE detector experiment operates in Frascati, at the DA$\phi$NE $\phi$-factory. It consists of three main parts, a cylindrical drift chamber (DC)~\cite{DC} surrounded by an electromagnetic calorimeter (EMC)~\cite{EMC}, all embedded in a magnetic field of 0.52 T, provided along the beam axis by a superconducting coil located around the calorimeter. The EMC energy and time resolutions are $\sigma_E/E = 5.7\%/\sqrt{E\text{[GeV]}}$ and $\sigma_t(E)=57\text{ps}/\sqrt{E\text{[GeV]}}\oplus100 \text{ps}$, respectively. The EMC consist of a barrel and two end-caps of lead/scintillating fibers, which cover $98\%$ of the solid angle. The all-stereo drift chamber, $4\text{m}$ in diameter and $3.3\text{m}$ long, operates with a light gas mixture ($90\%$ helium, $10\%$ isobutane). The position resolutions are $\sigma_{xy} \sim 150 \mu\text{m}$ and $\sigma_z \sim 2\text{mm}$. Momentum resolution, $\sigma_{p\perp}/p_{\perp}$, is better than $0.4\%$ for large angle tracks.
 
\section{U boson search in $\phi \rightarrow \eta U$ with $U \rightarrow e^+e^-$}
The first search of the U boson at KLOE was the decay $U \rightarrow e^+e^-$ in the process $\phi \rightarrow \eta U$. From a sample of $1.5 \,\text{fb}^{-1}$ of data collected during the 2004-2005 data taking, a total of 13000 events of $\eta \rightarrow \pi^+ \pi^- \pi^0$ with an associated $e^+e^-$ pair were selected. In a second analysis, a data sample of 31000 events of $\eta \rightarrow \pi^0\pi^0\pi^0$ with an associated $e^+e^-$ pair were selected from a $1.7 \,\text{fb}^{-1}$ of data from 2004-2005. The corresponding background contributions were of the order of $\sim 2\%$~\cite{eta1} and $\sim 3\%$~\cite{eta2}, respectively.
The irreducible background from the Dalitz decay $\phi \rightarrow \eta \gamma^* \rightarrow \eta e^+e^-$ was directly extracted from the data by a fit to the $M_{ee}$ distribution parameterized according to the Vector Meson Dominance model~\cite{vmd}. 

\begin{figure}[htb!]
\centerline{%
\includegraphics[width=6.5cm]{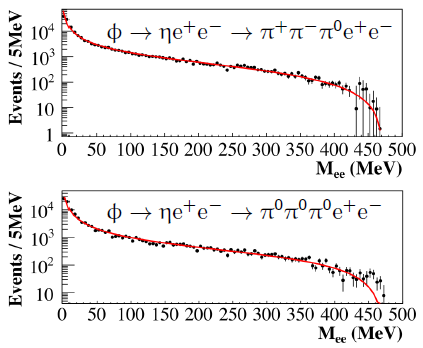}}
\caption{Di-electron invariant mass distributions, $\text{M}_{ee}$, for $\phi \rightarrow \eta e^+e^-$ with $\eta \rightarrow \pi^+\pi^-\pi^0$ ({\bf{top}}) and $\eta \rightarrow \pi^0\pi^0\pi^0$ ({\bf{bottom}}). The red lines are the fits to the measured data.}
\label{fig:fig1}
\end{figure}

As can be seen in Fig.~\ref{fig:fig1}, no resonant signal is observed in the $M_{ee}$ distributions of both analyses. While the peak around $400 \,\text{MeV}/\text{c}^2$ is due to background from the decay $\phi \rightarrow K_SK_L$. The Confidence Levels (CLs) technique~\cite{cls} was used to set an upper limit on the kinetic mixing parameter, as a function of the U boson mass, using the signal cross section given by~\cite{wang},

\begin{equation}
\sigma(\phi \rightarrow \eta U) \sim \epsilon^2 |F_{\eta\phi}(m_U^2)|^2\sigma(\phi \rightarrow \eta\gamma)
\end{equation}
The $90\%$ confidence level limit is presented in Fig.~\ref{fig:fig_results}

\section{U boson search in $e^+e^- \rightarrow U \gamma$ with $U \rightarrow \mu^+\mu^-$}
The U boson was also searched in the process $e^+e^- \rightarrow U \gamma$ with $U \rightarrow \mu^+\mu^-$, in a sample of $239.3 \,\text{pb}^{-1}$ of data collected in 2002~\cite{mumu}. The expected signal would show up as a narrow resonance in the di-muon mass spectrum.

The candidate events were selected by requiring two opposite charged tracks emitted at large polar angles, with an initial-state radiation (ISR) photon emitted at small angles, and thus undetected. The photon was later kinematically reconstructed from the charged leptons.

\begin{figure}[htb!]
\centerline{%
\includegraphics[width=6.5cm]{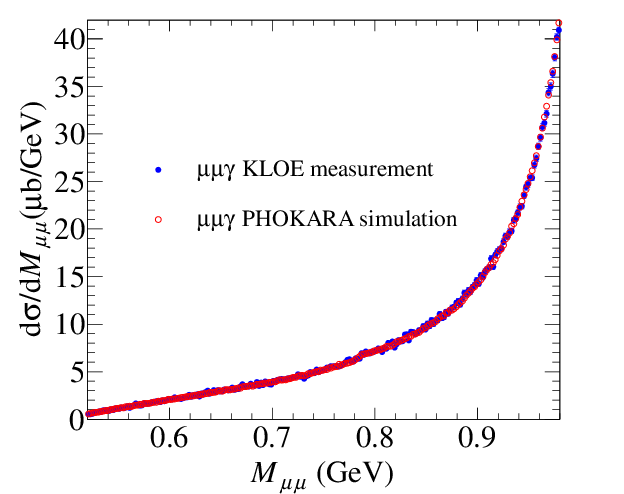}}
\caption{Di-muon invariant mass distributions, $\text{M}_{\mu\mu}$. Comparison of data (full blue circles) and simulation (open red circles).}
\label{fig:fig2}
\end{figure}

Using energy and momentum conservation, a variable called {\it{``track mass''}}, $M_{trk}$ was used to separate muons from pions and electrons. The $M_{trk}$ was calculated assuming two opposite charged tracks of equal mass and an unobserved photon in the final-state.

Residual backgrounds were determined using Monte Carlo simulation by fitting the observed $M_{trk}$ spectrum. The resulting invariant mass spectrum was obtained after subtracting residual backgrounds and dividing by efficiency and luminosity. Figure~\ref{fig:fig2} shows the di-muon invariant mass, which is in excellent agreement with the PHOKARA Monte Carlo simulation.
Since no resonant peak was observed, the CLs technique was used to estimate the number of U boson signal events excluded at $90\%$ confidence level, $N_{CLs}$ and then the limit on the kinetic mixing parameter,

\begin{equation}
\epsilon^2 = {\alpha_{D} \over \alpha_{EM}} = {N_{CLs} \over \epsilon_{eff}}{1 \over H \cdot I \cdot L_{integrated}}
\label{eq:radiator}
\end{equation}

where $\epsilon_{eff}$ is the overall efficiency, $I$ is the effective cross section, $L_{integrated}$ the integrated luminosity and $H$ is the radiator function, which is extracted from the differential cross section, $d\sigma_{\mu\mu\gamma}/dM_{\mu\mu}$. A systematic uncertainty of about $2\%$ was estimated. The $90\%$ confidence level limit is shown in Fig.~\ref{fig:fig_results}

\section{U boson search in $e^+e^- \rightarrow U \gamma$ with $U \rightarrow e^+e^-$}

The study of the reaction $e^+e^- \rightarrow U \gamma$, $U \rightarrow e^+e^-$, is similar to the previously described analysis but with the characteristic that allows to investigate the low mass region close to the di-electron mass threshold~\cite{anthony}. 

\begin{figure}[htb!]
\centerline{%
\includegraphics[width=6.5cm]{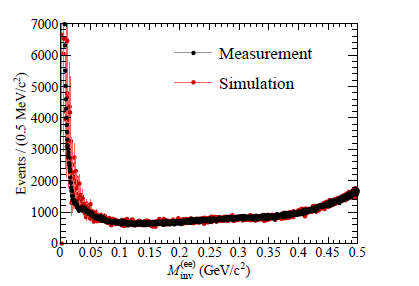}}
\caption{Di-electron invariant mass distribution, $M_{ee}$, for the process $e^+e^- \rightarrow e^+e^- \gamma$ (black circles) compared to the MC simulated spectra (red circles).}
\label{fig:fig3}
\end{figure}

\begin{figure}[htb!]
\centerline{%
\includegraphics[width=8.5cm]{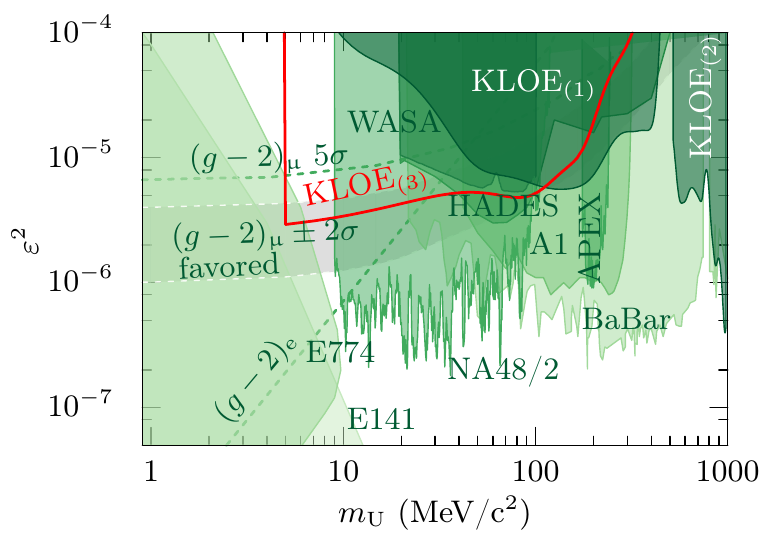}}
\caption{Exclusion limits on the kinetic mixing parameter, $\epsilon^2$, from KLOE (in red): $\text{KLOE}_1$, $\text{KLOE}_2$ and $\text{KLOE}_3$ correspond to the combined limits from the analysis of $\phi \rightarrow \eta e^+e^-$, $e^+e^- \rightarrow \mu^+\mu^-\gamma$ and $e^+e^- \rightarrow e^+e^-\gamma$, respectively. The results are compared with the limits from E141, E774~\cite{e141}, MAMI/A1~\cite{mami}, APEX~\cite{apex}, WASA~\cite{wasa}, HADES~\cite{hades}, NA48/2~\cite{na48} and BaBar~\cite{babar_ress}. The gray band indicates the parameter space favored by the ($\text{g}_{\mu}-2$) discrepancy.}
\label{fig:fig_results}
\end{figure}

For the event selection, two opposite charged tracks and a photon  were required. To reduce the background contamination a pseudo-likelihood discriminant was used to separate electrons from muons and pions, and then the {\it{"track mass"}} variable, $M_{trk}$, was also used to further discriminate the background sources. The resulting background contamination was less than $1.5\%$. The Fig.~\ref{fig:fig3} compares the di-electron invariant mass to MC BABAYAGA-NLO simulation~\cite{babayaga} modified to allow the Bhabha radiative process to proceed only via the annihilation channel, in which the U boson signal would occur, showing an excellent agreement.  

The upper limit of the kinetic mixing parameter as a function of $m_U$ was evaluated with the CLs technique in an analogous way as the $e^+e^- \rightarrow \mu^+\mu^-\gamma$. The limit on the U boson signal was evaluated at $90\%$ confidence level and the limit in the kinetic parameter was calculated using equation~(\ref{eq:radiator}). In this case the selection efficiency amounts to $\epsilon_{eff} \sim 1.5-2.5\%$ and the integrated luminosity corresponds to $L_{integrated} = 1.54 \,\text{fb}^{-1}$ from the 2004-2005 data campaign.

\section{U boson search in $e^+e^- \rightarrow h'U$ with $U \rightarrow \mu^+\mu^-$}
A natural consequence of the mass of the U boson is the breaking of the $U_D$ hidden symmetry associated by a Higgs-like mechanism through an additional scalar particle, called $h'$ or dark Higgs. The production cross section of the dark Higgstrahlung process, $e^+e^- \rightarrow h'U$ with $U \rightarrow \mu^+\mu^-$, would be proportional to the product $a_{D}\times\epsilon^2$~\cite{higgs}. Thus this process is suppressed by a factor $\epsilon$ comparing to the previous processes, already suppressed by a factor $\epsilon^2$. Depending on the relative masses of the $h'$ and the U boson there are two possible decay scenarios: if $m_{h'} > 2m_U$, the dark Higgs could decay via $h' \rightarrow UU \rightarrow 4l, 4\pi, 2l + 2\pi$, where $l$ denotes lepton. This scenario was studied by Babar~\cite{babar} and Belle~\cite{belle} in recent experiments. If $m_{h'} < 2m_U$, then the dark Higgs would have a large lifetime and would escape any detection. This "invisible" dark Higgs scenario has been the object of study by KLOE.

\begin{figure}[htb!]
\centerline{%
\includegraphics[width=5.5cm]{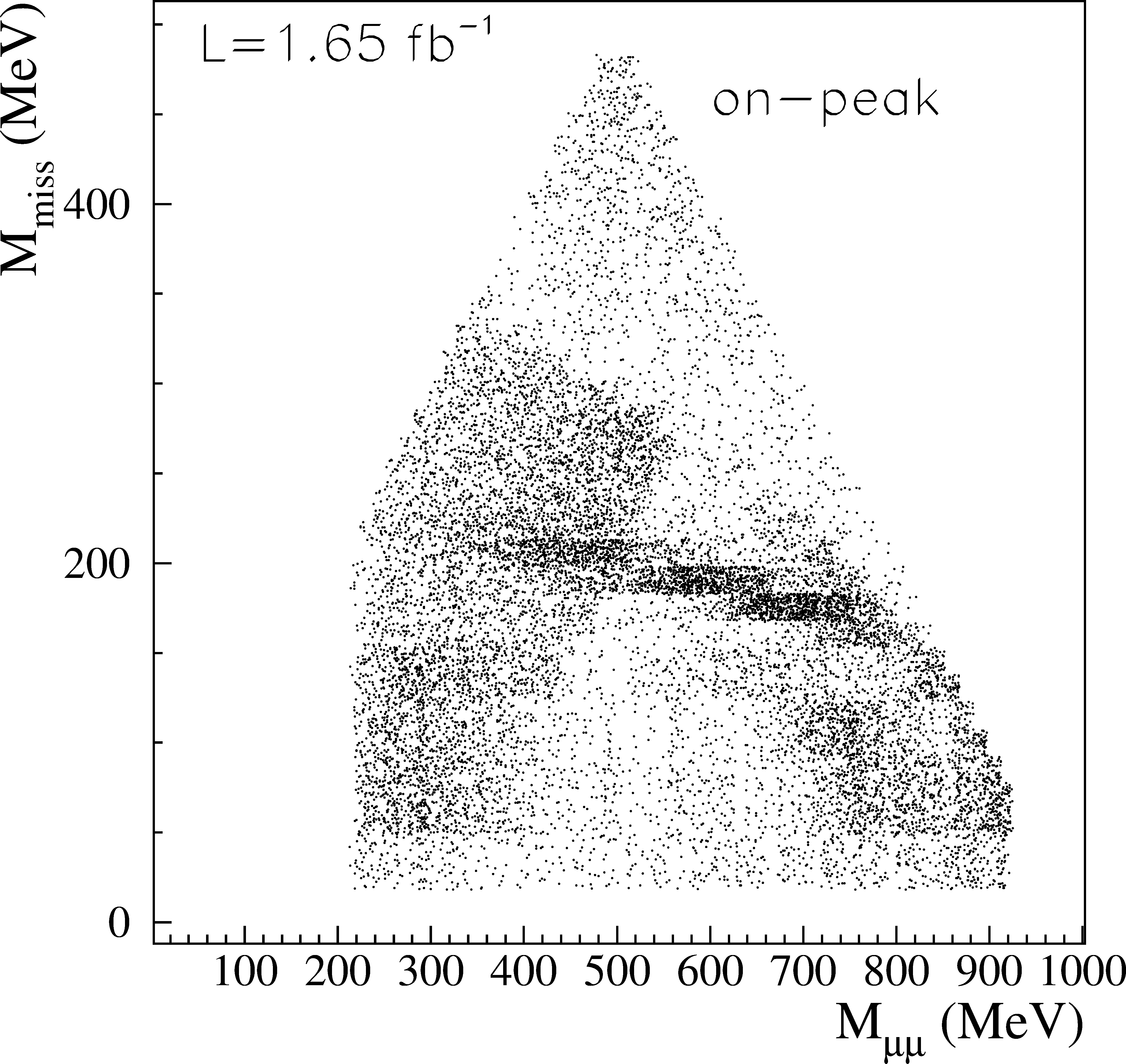}
\includegraphics[width=5.5cm]{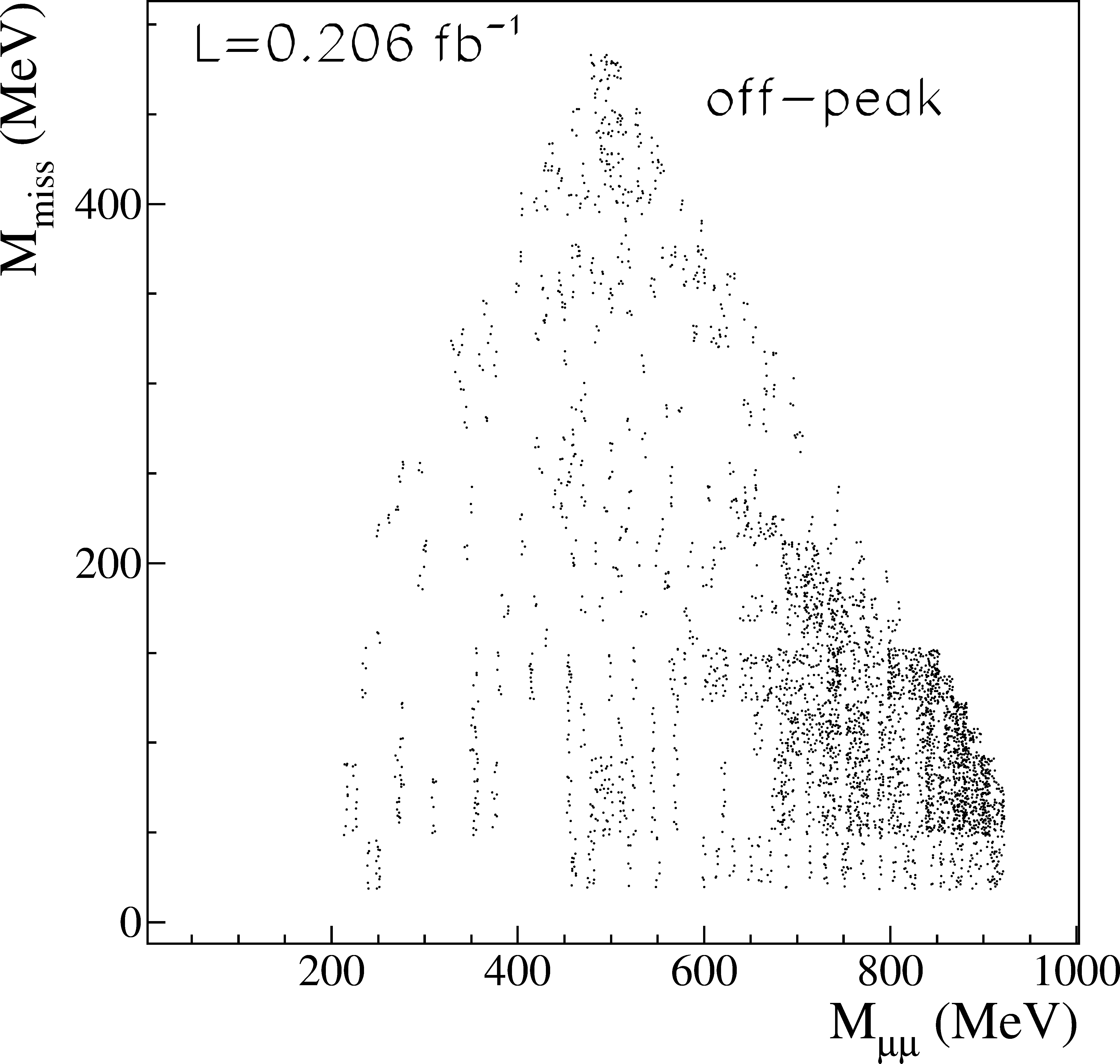}}
\caption{Missing mass, $M_{miss}$, versus di-muon mass, $M_{\mu\mu}$, for the $1.65\,\text{fb}^{-1}$ on-peak data sample ({\bf{left}}) and the $0.2 \,\text{fb}^{-1}$ off-peak sample ({\bf{right}}).}
\label{fig:fig4}
\end{figure}

The analysis was performed on $1.65 \,\text{fb}^{-1}$ of data collected during 2004-2005 data campaign at a center of mass energy at the $\phi$-peak and on a data sample of $0.2 \,\text{fb}^{-1}$ at a center of mass energy of $\sim 1000 \,\text{MeV}$. The expected signal would show up as a sharp enhancement in the missing mass, $M_{miss}$, versus $\mu\mu$ invariant mass, $M_{\mu\mu}$, two-dimensional spectra~\cite{dark_higgstrahlung}, shown in Fig.~\ref{fig:fig4}. 

Since most of the signal is expected to be in just one bin, a sliding matrix of $5 \times 5$ bins was built and used with data and Monte Carlo to check the presence of a possible signal in the central bin while the neighboring cells were used to estimate the background. The evaluated selection efficiencies were found to be about $15\%-25\%$.

\begin{figure}[htb!]
\centerline{%
\includegraphics[width=5.5cm]{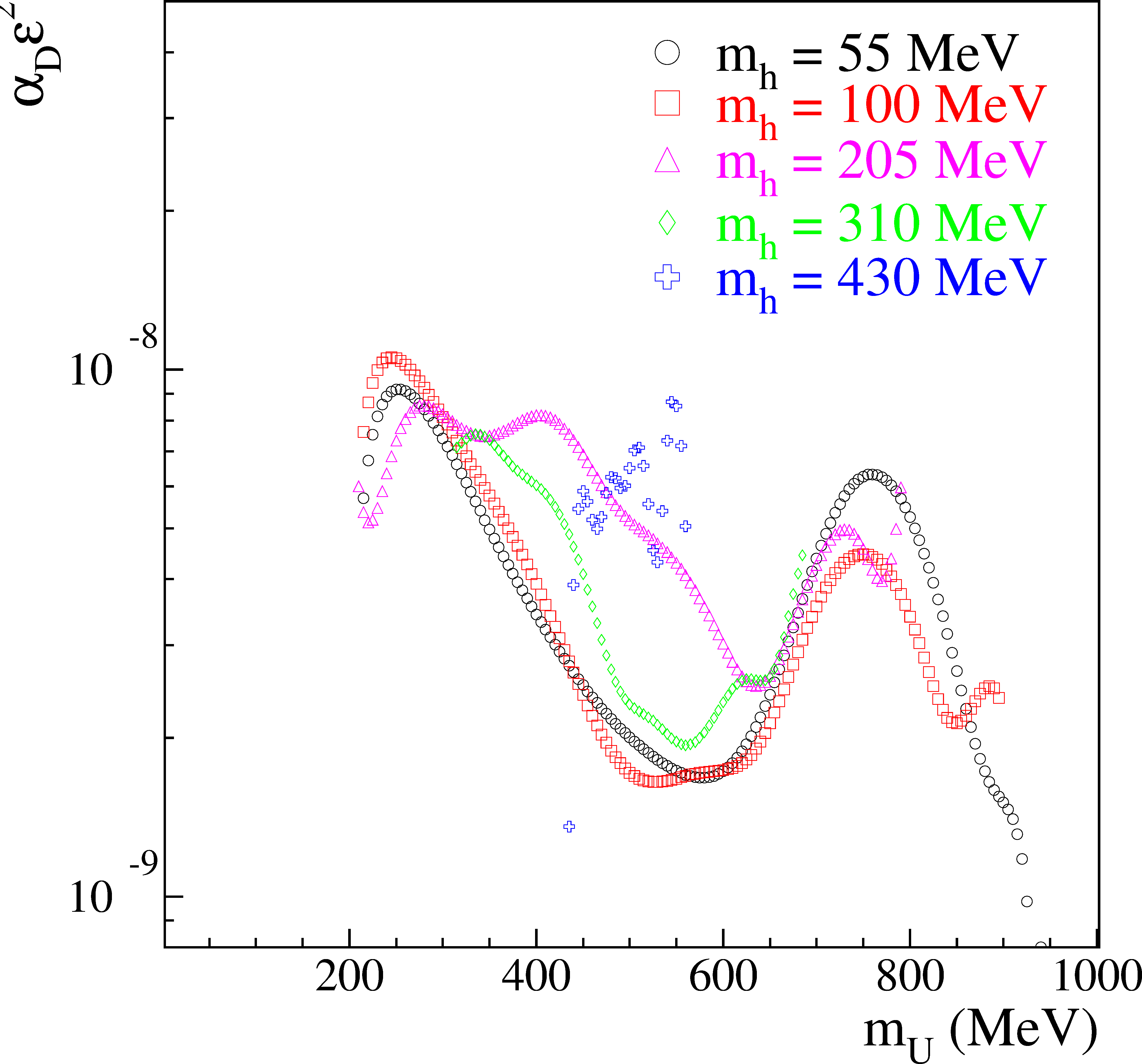}
\includegraphics[width=5.5cm]{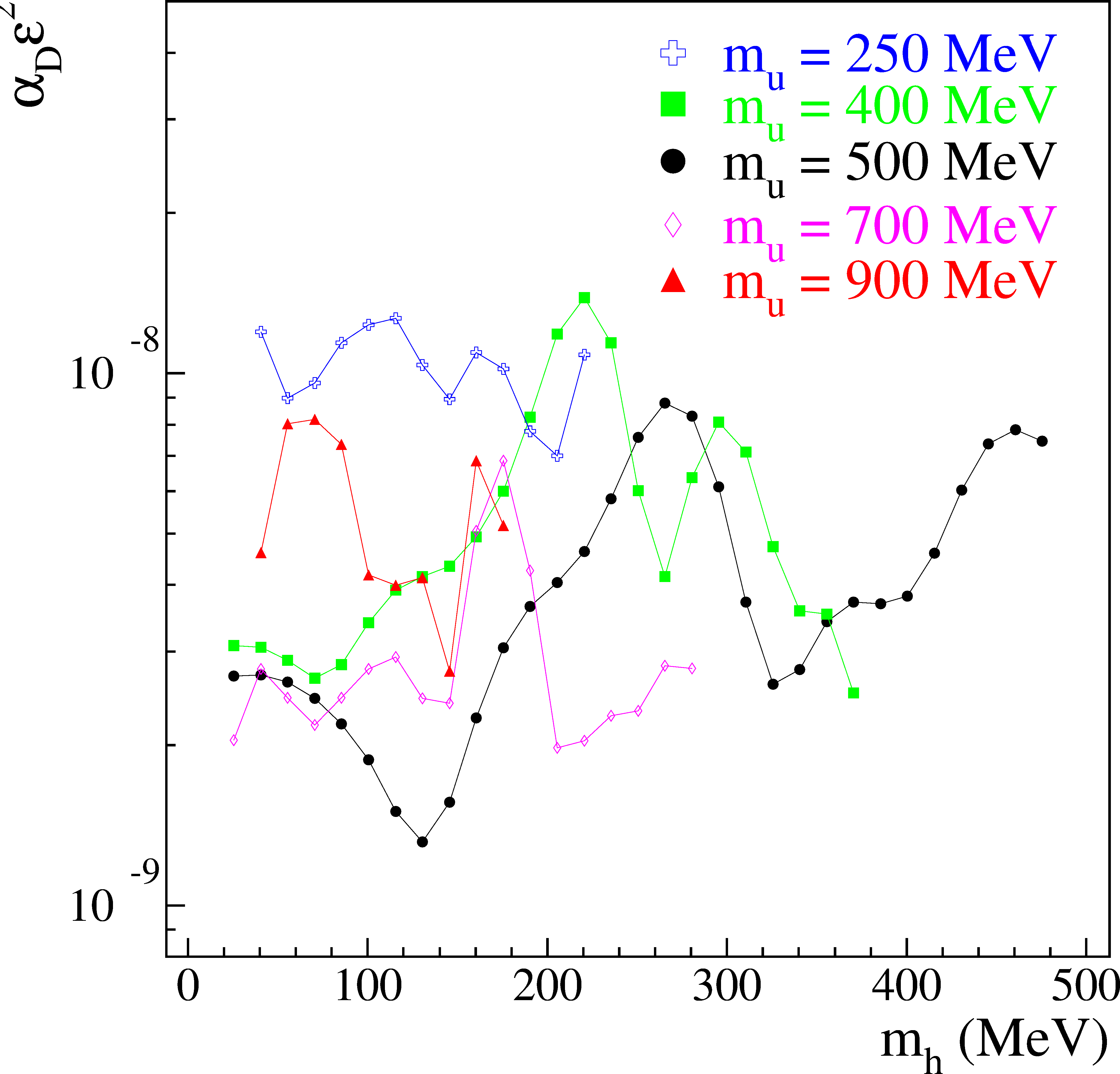}}
\caption{Combined $90\%$ confidence level upper limits in $\alpha_D \times \epsilon^2$ as a function of $m_U$ for different $m_{h'}$ values ({\bf{left}}) and as a function of $m_{h'}$ for different $m_U$ ({\bf{right}}).}
\label{fig:fig5}
\end{figure}

The different sources of background can be identified in Fig.~\ref{fig:fig4}, with its different contributions from $\phi \rightarrow K^+K^-, \, K^{\pm} \rightarrow \mu^{\pm}\nu$, $\phi \rightarrow \pi^+\pi^-\pi^0$, $e^+e^- \rightarrow \mu^+\mu^-, \, \pi^+\pi^-$, $e^+e^- \rightarrow e^+e^-\mu^+\mu^-$ and $e^+e^- \rightarrow e^+e^-\pi^+\pi^-$. In the right plot of Fig.~\ref{fig:fig4} (off-peak sample), all the backgrounds from the $\phi$ decays are strongly suppressed.
No signal of the dark Higgstrahlung process was observed and a Bayesian limit on the number of signal events, $N_{90\%}$, was derived for both samples separately. The product $\alpha_D \times \epsilon^2$ was then calculated according to,

\begin{equation}
\alpha_D \times \epsilon^2 = {N_{90\%} \over \epsilon_{eff}}{1 \over \sigma_{h'U}(\alpha_D\epsilon^2 = 1) \cdot L_{integrated}}
\end{equation}
with,
\begin{equation}
\sigma_{h'U} \propto {1 \over s}{1 \over (1-m_{U}^2/s)^2}
\end{equation}

and where $\alpha_D \times \epsilon^2$ is assumed to be equal 1. A conservative $10\%$ of systematic uncertainty was considered. The combined $90\%$ confidence level limits for both on- and off-peak data samples are presented in Fig.~\ref{fig:fig5}, as a function of $m_U$ (left) and of $m_{h'}$ (right). The limit values of $\alpha_D \times \epsilon^2$ of $10^{-9} - 10^{-8}$ at $90\%$ confidence level translate into a limit on the kinetic parameter, $\epsilon^2$, of $10^{-6} - 10^{-8}$ ($\alpha_D = \alpha_{EM}$). 

\section{Conclusions}
The KLOE collaboration has extensively contributed to the U boson searches by analyzing four different production processes. Up to now, no evidence for a U boson or dark Higgs boson was found and limits at the $90\%$ confidence level were set on the kinetic mixing parameter, $\epsilon$, in the mass range $5 \,\text{MeV} < m_U < 980 \,\text{MeV}$. Also, limits on $\alpha_D \times \epsilon^2$ at the $90\%$ confidence level in the parameter space $2m_{\mu} <  m_U < 1000 \,\text{MeV}$ with $m_{h'} < m_U$ have been extracted from the search for the U boson in the dark Higgstrahlung process. In the meantime a new data campaign has started with the KLOE-2 setup, which will collect more than $5 \,\text{fb}^{-1}$ in the next three years. The new setup and the enlarged statistics could further improve the current limits on the dark coupling constant by at least a factor of two.

We warmly thank our former KLOE colleagues for the access to the data collected during the KLOE data taking campaign.
We thank the DA$\Phi$NE team for their efforts in maintaining low background running conditions and their collaboration during all data taking. We want to thank our technical staff: 
G.F. Fortugno and F. Sborzacchi for their dedication in ensuring efficient operation of the KLOE computing facilities; 
M. Anelli for his continuous attention to the gas system and detector safety; 
A. Balla, M. Gatta, G. Corradi and G. Papalino for electronics maintenance; 
M. Santoni, G. Paoluzzi and R. Rosellini for general detector support; 
C. Piscitelli for his help during major maintenance periods. 
This work was supported in part by the EU Integrated Infrastructure Initiative Hadron Physics Project under contract number RII3-CT- 2004-506078; by the European Commission under the 7th Framework Programme through the `Research Infrastructures' action of the `Capacities' Programme, Call: FP7-INFRASTRUCTURES-2008-1, Grant Agreement No. 227431; by the Polish National Science Centre through the Grants No.\
2011/03/N/ST2/02652,
2013/08/M/ST2/00323,
2013/11/B/ST2/04245,
2014/14/E/ST2/00262,
2014/12/S/ST2/00459.

\end{document}